\documentclass[12pt,epsf]{article}
\usepackage{amsmath}
\usepackage{amssymb}
\usepackage{color}
\usepackage{axodraw}
\usepackage{graphics}
\input epsf

\begin{document}
\baselineskip 24pt

\newcommand{\be}{\begin{equation}}
\newcommand{\ee}{\end{equation}}

\newcommand{\bea}{\begin{eqnarray}}
\newcommand{\eea}{\end{eqnarray}}
\newcommand{\nn}{\nonumber}
\newcommand{\eq}{\ref}
\newcommand{\ol}{\overline}

\newcommand{\sheptitle}
{Inflationary Solution to the Strong CP and $\mu$ Problems}

\newcommand{\shepauthor}
{O. J. Eyton-Williams and S. F. King}

\newcommand{\shepaddress}
{School of Physics and Astronomy, University of Southampton, \\
        Southampton, SO17 1BJ, UK}
\vspace{0.25in}


\newcommand{\shepabstract} {We show that the vacuum expectation 
value of the inflaton at the Peccei-Quinn axion scale can generate
the supersymmetric Higgs mass $\mu$ term. This provides an inflationary simultaneous
solution to the strong CP problem and the $\mu$ problem of the Minimal
Supersymmetric Standard Model,
and gives a testable prediction for the $\mu$ parameter:
$\mu^2 \approx (0.25-0.5)m_0^2$, where $m_0$ is the soft Higgs scalar mass. 
Our model involves a very small Yukawa coupling of order 
$10^{-10}$, which could originate from 
an extra-dimensional scenario or type I string theory.}

\begin{titlepage}
\begin{flushright}
hep-ph/0411170 \\
\end{flushright}
\begin{center}
{\large{\bf \sheptitle}}
\\ \shepauthor \\ \mbox{} \\ {\it \shepaddress} \\
{\bf Abstract} \bigskip \end{center} \setcounter{page}{0}
\shepabstract
\begin{flushleft}
\today
\end{flushleft}

\vskip 0.1in
\noindent

\end{titlepage}

\newpage

\section{Introduction} \label{sec:Introduction}

The $\mu$ problem of the Minimal Supersymmetric Standard Model
(MSSM), the origin of the Supersymmetric Higgs mass parameter
$\mu H_uH_d$ where $H_u,H_d$ are the two Higgs doublets and
$\mu$ is of the same order of magnitude as the soft supersymmetric (SUSY) breaking parameters,
has long been a puzzle \cite{Chung:2003fi}.
Another puzzle is the physical nature of the scalar field
which drives cosmological inflation, known as the inflaton
field. It is well known that the inflaton cannot be identified
with the Higgs fields of either the Standard Model or one
of its SUSY extensions, and there are few physical 
candidates for the inflaton field in the literature \cite{Lyth:1998xn}.

The possible connection between the strong CP problem and the $\mu$
problem in supersymmetry was explored some time ago \cite{Nilles:1981py}, 
and a non-renormalisable operator responsible for generating the 
$\mu$ term was proposed in \cite{Gherghetta:1995jx}.
The first simultaneous solution to the strong CP problem
and $\mu$ problem based on {\em renormalisable} operators
was proposed in \cite{Bastero-Gil:1997vn}.
In \cite{Bastero-Gil:1997vn} the $\mu$ term is
generated by the VEV of a singlet field $N$, 
in a similar way to the Next-to-Minimal Supersymmetric Standard Model
(NMSSM) \cite{NMSSM, NMSSMphenom}:
$\lambda N H_uH_d\rightarrow \mu H_uH_d$, where $\mu = \lambda <N>$.  
However, whereas in the NMSSM the vacuum expectation value (VEV)
of the singlet field $N$ takes a value of order the electroweak
breaking scale, in \cite{Bastero-Gil:1997vn} its VEV
is of order the Peccei-Quinn symmetry breaking scale
\cite{Peccei:1977hh}, allowing an invisible axion
solution to the strong CP problem \cite{Zhitnitsky:1980tq, Dine:1981rt}.
Since the $\mu$ parameter
must be of order the TeV scale, this implies that the
dimensionless Yukawa coupling $\lambda$ must be extremely small,
possibly of order $10^{-10}$ \cite{Bastero-Gil:1997vn}.

The scenario proposed in \cite{Bastero-Gil:1997vn} also
provides a model of inflation
since the NMSSM operator $\kappa N^3$ is replaced by the
operator $\kappa \phi N^2$,
where $\phi$ is identified as the inflaton field
and $N$ as the waterfall field of hybrid inflation
\cite{Bastero-Gil:1997vn}. Whereas the NMSSM operator $\kappa N^3$
is responsible for a $\mathbb{Z}_3$ symmetry, leading to problems with
cosmological domain walls when it breaks,  
the term $\kappa \phi N^2$ permits a global $U(1)_{PQ}$
symmetry, leading to a solution to the strong CP problem 
\cite{Bastero-Gil:1997vn}.
It also allows hybrid inflation providing the 
dimensionless Yukawa couplings satisfy
$\lambda \sim \kappa \sim 10^{-10}$. Such small Yukawa couplings could
arise from an extra-dimensional scenario 
due to volume suppression \cite{Bastero-Gil:2004ae}.
Note that the presence of the term $\kappa \phi N^2$ is
crucial not only to allow hybrid inflation to proceed but also 
to stabilise the potential in a natural way.
\footnote{Models with only the term $\lambda N H_uH_d$ have also subsequently been considered \cite{Miller:2003hm}, but without the additional term
$\kappa \phi N^2$ the vacuum is not necessarily stable. S.K. is grateful
to R. Nevzorov for pointing this out.}

In this paper we discuss a model in which 
the $\mu$ term is provided by the
same inflaton field which drives the super-luminal expansion
of the early universe. To be precise,
we suggest a simultaneous solution to the 
strong CP and $\mu$ problems in the framework of hybrid inflation
in which the $\mu$ term is generated by an operator
$\lambda \phi H_uH_d$ where $\phi$ is the inflaton field.
The $\mu$ term is then generated by the VEV of the inflaton field $\phi$
at the end of inflation:
$\lambda \phi H_uH_d\rightarrow \mu H_uH_d$ 
where $\mu = \lambda <\phi>$.  
We shall also require a term $\kappa \phi N^2$ which is crucial
to maintain the stability of the potential, where $N$ still plays
the part of the waterfall field in hybrid inflation. 
The above variation is interesting since, unlike the original
version of the model, it 
leads to a testable prediction of the
$\mu$ parameter: $\mu^2 \approx (0.25-0.5)m_0^2$, where $m_0$ is the soft Higgs scalar mass. \footnote{This soft mass is assumed to be universal for both $H_u$, $H_d$ and the $N$ field.  This universality is a feature of the model's type I string construction, derived in \cite{companion}.}
The generation of the $\mu$ term by the inflaton field also
implies deeper connections between SUSY Higgs phenomenology,
inflation, and the strong CP problem, and from a theoretical point of 
view admits a type I string theory embedding \cite{companion}.

We shall first outline the particle content and interactions of our
model. Then, in section \ref{sec:MinSoft} 
we discuss the potential and the minimum reached at
the end of inflation.  To stabilise this minimum and end inflation we
must require that the ratio of the soft mass and the trilinear falls
within a certain range which leads to the above prediction for the
$\mu$ parameter. Then, in section \ref{eq:Inflation}, we review some basic
inflationary requirements. Section \ref{sec:Conclusions} concludes the paper.

\section{The Model} \label{sec:TheModel}

To begin we define the model in terms of a superpotential and the
soft potential:
\begin{align}
  W=\lambda \phi H_u H_d + \kappa \phi N^2 \label{eq:NewW}
\end{align}
\begin{align}
  V_{soft} = \ &  V(0) + \lambda A_\lambda \phi H_u H_d + \kappa A_\kappa \phi N^2 + h.c. + \nonumber \\ 
   & m_0^2\left(|N|^2 + |H_u|^2 + |H_d|^2\right) - m_{\phi}^2|\phi|^2. \label{eq:NewV}
\end{align}
Here $\phi$ and $N$ are, respectively, the
inflaton and waterfall fields and are singlets of the
MSSM gauge group responsible for inflation. 
The Higgs fields $H_u,H_d$ have standard MSSM quantum
numbers. The dimensionless couplings $\lambda , \kappa$ are
$\mathcal{O}(10^{-10})$, and we have assumed a common scalar soft mass
squared for the Higgs and $N$ fields, but allowed 
a different (lighter) negative, soft mass squared for the inflaton
field $\phi$ in order to satisfy the slow roll conditions and yield an acceptable inflationary trajectory.

The generation of
the $\mu$ term is similar to that of the NMSSM, but the NMSSM is plagued by domain walls
\cite{Kibble:1976sj, Zeldovich:1974uw, Abel:1995wk, Vilenkin:1984ib}
(associated with breaking a discrete symmetry) created in the early
universe.  Our model does not face this problem since it does not have
an $N^3$ term and therefore replaces the discrete $\mathbb{Z}_3$ symmetry with
the continuous PQ symmetry mentioned above.  The PQ domain wall problem is discussed in section \ref{eq:Inflation}.
The charges of $\phi$, $N$ and the Higgs under the PQ symmetry must satisfy the following requirements
\begin{equation}
  Q_\phi + Q_{H_u} + Q_{H_d} = 0, \ \ Q_\phi + 2 Q_N = 0
\end{equation}
and the quark fields have the usual axial PQ charges.

\section{The Potential} \label{sec:MinSoft}

In this section we construct and minimise the potential and calculate
the VEVs relevant to our model. We initially search the potential in
the region of zero Higgs VEV post inflation.  For our model to map
on to the MSSM at low energies the Higgs must be minimised at zero at
high scales. Subsequently radiative electroweak symmetry breaking
(EWSB) then occurs in the usual way, resulting in non-zero Higgs
VEVs at low energy. We shall not discuss this radiative EWSB mechanism
further in this paper, since it is well known, but instead shall
confine our attention to showing that the Higgs VEVs are indeed
zero at high energy. Thus the VEV of the inflaton generates an
effective TeV scale $\mu$ term, leading to an effective MSSM theory
valid below the PQ scale with standard EWSB.

For the first step in the derivation we write down the relevant parts
 of the supersymmetric scalar potential (derived from the
 superpotential Eq.\,\ref{eq:NewW}) and the soft scalar potential:
\begin{align}
  V_{susy}= &|\lambda H_u H_d + \kappa N^2|^2 + \lambda^2|\phi H_u|^2 + \lambda^2|\phi H_d|^2 + 4\kappa^2|\phi N|^2, \label{eq:VSusy} \\
  V_{soft}= & V(0) + \lambda A_\lambda\phi H_u H_d + \kappa A_\kappa\phi N^2 + h.c. \\ 
  & m_0^2\left(|H_u|^2 + |H_d|^2 + |N|^2\right) - m^2_\phi |\phi|^2.
\end{align}
The full scalar potential is given by $V= V_{susy}+V_{soft}$.
Henceforth, for this section, we set $\lambda=\kappa$, $A_\lambda=A_\kappa$. This is done for simplicity here, 
but can be justified in terms of an explicit high scale type I string model.

Since the Higgs fields will eventually achieve TeV scale VEVs,
whereas the $N$ and $\phi$ fields achieve PQ scale VEVs, their contribution to the
energy density will be quite negligible. \footnote{Note that this approximation is not valid for the models in
  \cite{Miller:2003hm}.} Of course one must check
that the Higgses do not also receive PQ scale VEVs, and that their zero
tree-level VEVs represent a stable vacuum, which we will subsequently do.
Minimising the tree level potential gives:
\begin{align}
  <\phi> &= -\frac{A_\lambda}{4\lambda} \label{eq:Phi}
\end{align}
\begin{align}
  <N> &= \pm \frac{A_\lambda}{2\sqrt{2} \lambda}\sqrt{1 - \frac{4m_0^2}{A_\lambda^2}} \label{eq:N} 
\end{align}
\begin{align}
  <H_u>=<H_d>=0 \label{eq:H}
\end{align}
 where we have assumed that $m_\phi\approx 0$.  We will refer to this as the ``good'' minimum as it is phenomenologically preferred.

Looking back at Eq.\,\ref{eq:NewW} we see that when $\phi$ moves to its VEV we obtain a supersymmetric mass term for the Higgses, a $\mu$ term:
\begin{equation}
  \mu = -\lambda \frac{A_\lambda}{4 \lambda} = -\frac{A_\lambda}{4}. \label{eq:mu}
\end{equation}
Since $\lambda$ is the only dimensionless coupling in Eq.\,\ref{eq:VSusy} $\mu$ automatically appears at the electroweak scale.

The soft mass parameters are constrained by inflationary
requirements, and this will lead to the prediction of the 
$\mu$ parameter in our approach. 
The requirement that inflation ends implies
$A_\lambda^2 > 4m_0^2$ as a necessary condition.  If $A_\lambda^2 \leq 4
m_0^2$ then $N$ only has a minimum at zero and never destabilises to end
inflation.  In our model we have this bound and an additional upper
bound on the trilinears which we will now derive.

Now we need to show that the ``good'' solution is a minimum of the potential (in the absence
of radiative corrections).
It is important to check that $<H_{u/d}>\,=0$ since
we do not want electroweak symmetry to be
broken at the high scale. In order to check this we first need to locate the turning points to ensure that $H_u=H_d=0$ is a valid solution.  Then we must examine this point to see if it is a minimum.

Solving $\frac{\partial V}{\partial H_u}=0$ for $H_u$ gives us turning points for $H_u$ and, since the potential is
symmetric under interchange of $H_u$ and $H_d$, the solutions to
$\frac{\partial V}{\partial H_d}=0$ and $\frac{\partial V}{\partial
H_u}=0$ must be related by exchanging $H_u$ and $H_d$.  As a result we
can solve $\frac{\partial V}{\partial H_u}=0$ by setting $H_u=H_d=H$.
We find two non-trivial solutions namely the ``good'' solution 
in Eqs.\ref{eq:Phi},\ref{eq:N},\ref{eq:H},
and another with $<H>\neq0$ which we will refer to as the ``bad'' solution on account of its unphysically large Higgs VEV:
\begin{align}
  <H>\,=\pm \frac{A_\lambda}{2\lambda}\sqrt{1-\frac{4m_0^2}{A_\lambda^2}}.
\end{align}
The discussion of the ``bad'' solution will be deferred until appendix \ref{sec:GlobalMin}.  We also note that there exists a trivial solution (a maximum) with all fields at zero.

Now that we have shown that $H=0$, and by extension the ``good'' solution, is valid we want to determine the conditions under which this solution is a local minimum of the potential.

To prove this we need to show that the
Hessian is positive definite.  If
\begin{align} 
  \left(\begin{array}{cccc}
      V_{H_u H_u} & V_{H_u H_d} & V_{H_u \phi} & V_{H_u N} \\
      V_{H_d H_u} & V_{H_d H_d} & V_{H_d \phi} & V_{H_d N} \\
      V_{\phi H_u} & V_{\phi H_d} & V_{\phi \phi} & V_{\phi N} \\
      V_{N H_u} & V_{N H_d} & V_{N \phi} & V_{N N}
    \end{array}\right) \label{eq:hessian}
\end{align}
is a positive definite matrix, then the ``good'' solution is a minimum.  To demonstrate this is true it is sufficient to show that all the eigenvalues of Eq.\,\ref{eq:hessian} are positive.  This requirement can be expressed in terms of the ratio between $|A_\lambda|$ and $m_0$ which we parametrise by $x=\frac{|A_\lambda|}{m_0}$.  We find that both $x^2>4$ and $x^2<8$ must be satisfied for the point to be a minimum.  Expressed as a function of the soft terms we have
\begin{align}
  8m_0^2>|A_\lambda|^2>4m_0^2. \label{eq:bounds}
\end{align}
For $|A_\lambda|^2>8m_0^2$ Eq.\,\ref{eq:hessian} has both positive and
negative eigenvalues and we would have a saddle point.

Since the $\mu$ parameter is given by Eq.\,\ref{eq:mu} the constraint in Eq.\,\ref{eq:bounds} leads to a prediction of the $\mu$ parameter in the range: \footnote{It should be pointed out at this stage that the ``good'' solution is not the global minimum of the potential.  The ramifications of this fact and potential solutions are discussed in appendix \ref{sec:GlobalMin}.}
\begin{align}
\mu^2=(0.25-0.5)m_0^2 \label{eq:muprediction}
\end{align}  

\section{Inflation} \label{eq:Inflation}
   Any model purporting to describe inflation must satisfy some basic
   requirements: it must have a field that is slowly rolling for a
   sufficient amount of expansion, it must predict curvature
   perturbations in line with CMB observations and its prediction for
   the spectral index must be consistent with current
   measurements. In particular it must satisfy the slow roll
   conditions, $\epsilon \ll 1$ and $\eta \ll 1$, and have a
   spectral index compatible with $n_s=0.99 \pm 0.04$
   \cite{Bunn:1996da, Bunn:1994ds}.
The two slow roll conditions are usually expressed as
  \begin{equation}
    \epsilon_N=\frac{1}{2}m_P^2\left(\frac{V'}{V}\right)^2 \ll 1
  \end{equation}
  \begin{equation}
   |\eta_N|=\left|m_P^2\frac{V''}{V}\right| \ll 1
  \end{equation}
  where $N$ specifies when, 
in terms of number of e-folds before the end of inflation
, $\epsilon$ and $\eta$ were evaluated.  They are evaluated at the time when the
  scales, that are currently just re-entering, left the horizon.  For our model,
  with its relatively small vacuum energy during inflation, $N\sim
  45$.  Here we are using $m_P=M_{\mbox{\tiny Planck}}/\sqrt{8\pi}$

  In hybrid inflation \cite{Linde:1990gz, Linde:1991km,
    Copeland:1994vg, Lyth:1996we, Linde:1997sj} during the
  inflationary epoch the inflaton field $\phi$ slowly rolls along some
  almost flat direction. A second ``waterfall'' field $N$ whose mass
  squared is positive during inflation, and hence whose field value is
  held at zero during inflation, is subsequently destabilised when the
  inflaton reaches a critical value. After this its mass squared
  becomes tachyonic and it rolls out to a non-zero value, effectively
  ending inflation. In fact, as is the case in our model, inverted
  hybrid inflation \cite{Lyth:1996kt} occurs if the soft mass squared
  for the inflaton is negative, and normal hybrid if the soft mass
  squared was positive.  In both cases there is a critical point that
  marks the transition from positive to negative effective mass
  squared for $N$.

  In the previous section we saw that there are two non-trivial minima that we labelled ``good'' and ``bad''.  Which minimum is reached depends on the inflationary trajectory.  If a critical point is reached at which $N$ destabilises first then the fields will fall into the ``good'' minimum.  On the other hand if the corresponding critical point for the Higgs is reached first then we roll out to the ``bad'' minimum.  It is therefore important to examine the critical points for the $H_u$, $H_d$ and $N$ fields.

  The critical values for the Higgs and $N$ fields can be derived from
  Eq.\,\ref{eq:hessian} by considering the stability of the Higgs and $N$ along a trajectory that has $\phi$ non-zero and all other fields set to zero.  The critical values of $\phi$ are roots of the eigenvalue
  equations in the Higgs and $N$ sectors and can be expressed in terms of the soft
  parameters. Clearly the $\phi$ sector is already unstable due to the
  negative soft mass squared for $\phi$.  In fact it has a positive gradient at this point: this is the origin of the slow roll.
  
  The critical points at which $N$ becomes unstable are
  \begin{equation}
    \phi_{\mbox{\tiny crit.(N)}} = \frac{A_\kappa}{4\kappa}\left(-1 \pm \sqrt{1-\frac{4m_0^2}{A_\kappa^2}}\right)
  \end{equation}
  and the Higgs fields destabilise at
  \begin{equation}
    \phi^-_{\mbox{\tiny crit.(H)}} = \frac{A_\lambda}{2\lambda}\left(-1\pm \sqrt{1-\frac{4m_0^2}{A_\lambda^2}}\right)
  \end{equation}
  and
  \begin{equation}
    \phi^+_{\mbox{\tiny crit.(H)}} = \frac{A_\lambda}{2\lambda}\left(1\pm \sqrt{1-\frac{4m_0^2}{A_\lambda^2}}\right)
  \end{equation}
  
  Within the ranges of $\phi$ bounded by these critical values the associated field is unstable. As a result our model requires an inverted hybrid inflationary trajectory that starts from a point with small, negative $\phi$ and all other fields held at zero by their positive effective masses. 

As $\phi$ rolls away from the origin it will reach $\phi_{\mbox{\tiny crit.(N)}}$ before $\phi^{-}_{\mbox{\tiny crit.(H)}}$, assuming that $m_0$ is non-zero, $\lambda=\kappa$ and $A_\lambda=A_\kappa$.  Therefore it the ``good'' minimum with $N \neq 0$ and $H_u=H_d=0$ that is reached on this trajectory.  We shall now discuss the slow roll period that occurs as $\phi$ moves away from the origin.

 For our trajectory, with all fields
  except the inflaton at zero, the potential simplifies to
  \begin{equation}
    V=V(0) - \frac{1}{2} m_\phi^2 \phi^2.
  \end{equation}
  In this case the slow roll conditions become
  \begin{equation}
    \epsilon_N=\frac{1}{2}\frac{m_P^2 m_\phi^4 \phi_N^2}{V(0)^2} \ll 1 \label{eq:epsN}
  \end{equation}
  \begin{equation}
    |\eta_N|=m_P^2\frac{|m_\phi^2|}{V(0)} \ll 1 \label{eq:eta}.
  \end{equation}
 Since
 \begin{equation}
   \phi_N=\phi_{\mbox{\tiny crit.(N)}} e^{N \eta} \label{eq:phiN}
 \end{equation}
 and $\eta \ll 1$ it follows that $\phi_N \sim \phi_{\mbox{\tiny crit.}}$. 
 Of course we must check that the slow roll conditions are satisfied.
 From Eq.\,\ref{eq:eta} we see that we have an upper limit on $m_\phi$
 of 10 MeV.
 However, from Eqs.\,\ref{eq:epsN} and \ref{eq:phiN} we require that,  $\eta_N < 0.25$, approximately.  If this were not enforced then $\phi_N$ would push $\epsilon_N$ above one. 
 This slightly lowers our upper limit on $m_\phi$ to 5 MeV. In our model $V(0)^{1/4} \sim 10^8$ GeV
 is fixed when we enforce zero vacuum energy at the minimum of the
 potential. This leads to a low Hubble constant during inflation of 
 $H\approx V(0)^{1/2}/3m_P\sim 1$ MeV and a low reheat temperature 
 after inflation.
 
 The reheat temperature is given by
\begin{align}
  T_{RH} \simeq 0.55g_*^{-1/4}\sqrt{\Gamma_\phi m_P}
\end{align}
 where \cite{Choi:1996vz} the decay rate is given by
\begin{align}
  \Gamma_\phi \sim \frac{M_\phi^3}{64\pi f_a^2}.
\end{align}
$M_\phi$ is the mass obtained after inflation and $f_a$ is the axion decay constant.  This simplifies to 
\begin{align}
  \Gamma_\phi \sim \frac{\lambda^2}{4\pi}M_\phi \sim 10^{-8}\mbox{ eV}
\end{align}
which leads to a reheat temperature of $T_{RH}\sim (1-10)$ GeV.
 The low reheat temperature slightly
 relaxes the upper bound on the axion decay constant, allowing
 $f_a\sim 10^{13}$ GeV \cite{Bastero-Gil:1997vn}.

 It turns out that the most stringent requirement on the masses comes
 from the density perturbation data.  From \cite{Liddle:1993fq} we see
 that 
 \begin{equation}
   \delta_H=\frac{32}{75}\frac{V(0)}{m_P^4}\epsilon_N^{-1} = 1.92 \times 10^{-5}.
 \end{equation}
 Satisfying this requirement with the inflaton would drive its mass
 down to below the eV scale.  This would require a high degree of
 fine-tuning. If the mass of the inflaton $\phi$ during inflation is
 in the MeV range this satisfies the slow roll constraints, but
 precludes the possibility that the density fluctuations are
 provided by the inflaton itself.  Thus extreme fine-tuning is
 alleviated \cite{Dimopoulos:2002kt} if we use a different field, a
 curvaton \cite{Lyth:2001nq, Moroi:2001ct, Moroi:2002rd}, to generate
 the curvature perturbations.  There are numerous examples of this mechanism 
 in the literature.  One possibility that might be compatible with our model
 is the axion as curvaton.  This case is explored in \cite{Dimopoulos:2003ii} though, at this stage, it is not clear whether this analysis is applicable to this model.
 Another possibility is to use the coupled curvaton mechanism
 \cite{Bastero-Gil:2002xr} in which the perturbations are provided by
 a second light scalar field which takes a non-zero value during
 inflation, and whose fluctuations are subsequently converted to
 curvature perturbations with the help of preheating effects.
 Alternatively we may appeal to a type of late-decaying curvaton
 mechanism which is consistent with low inflation scales with a
 symmetry breaking phase during inflation \cite{lyth}.

Tied into inflation is the issue of domain walls.  Since this model does not possess the $\mathbb{Z}_3$ symmetry of the NMSSM it sidesteps the domain wall problem encountered when $\mathbb{Z}_3$ breaks.  However, domain walls are also created when the PQ symmetry breaks \cite{Casas:1987gf, Casas:1987bw}.  During inflation the inflaton has a non-zero value hence breaks PQ symmetry spontaneously.  As a result the domain walls are created during inflation.  As such the exponential expansion of the universe will dilute them so that, by the end of inflation, their fraction of the total energy density will be negligible.

\section{Conclusions} \label{sec:Conclusions}

In this paper we have suggested that the field responsible for
cosmological inflation and the field responsible for generating the
$\mu$ term of the MSSM are one and the same.
We have shown that the vacuum expectation 
value of the inflaton at the Peccei-Quinn axion scale can generate
the supersymmetric Higgs mass $\mu$ term of the MSSM. This provides an inflationary simultaneous
solution to the strong CP problem and the $\mu$ problem of the MSSM,
and gives a testable prediction for the $\mu$ parameter:
$\mu^2 \approx (0.25-0.5)m_0^2$, where $m_0$ is the soft Higgs scalar mass. 
This implies deep connections between
supersymmetric Higgs phenomenology, inflation and the strong CP
problem.

Our model involves very small Yukawa couplings of order $10^{-10}$
which could originate from an extra-dimensional scenario
\cite{Bastero-Gil:2004ae}. In \cite{companion} we will 
show how such small Yukawa couplings can arise from 
embedding the model into type I string theory.
The string embedding will also post-justify the
assumptions that we have made here concerning smallness and equality
of the Yukawa couplings in Eqs.\,\ref{eq:NewW} and \ref{eq:NewV}, and
also the equality of the soft masses of the Higgses, $H_u$ and $H_d$, which we
have assumed to have the same soft mass as the $N$ field. 

Finally we note that Yukawa couplings as small as $10^{-10}$ 
allow the possibility of having
Dirac neutrino masses, which is testable in neutrino experiments and
would open up the possibility of relating the physics of the neutrino mass
scale to the physics of inflation, the strong CP problem and the
$\mu$ problem discussed here.

\begin{center} 
  \bf Acknowledgements
\end{center}
We would like to thank M. Bastero-Gil for helpful discussions,
and R. Nevzorov for reading the manuscript.
\appendix
  \section{Global Minima}\label{sec:GlobalMin}
In section \ref{sec:MinSoft} we discovered that ($<\phi>\,=-\frac{A_\lambda}{4\lambda}$, $<N>\,=\pm\frac{A_\lambda}{2\sqrt{2} \lambda}\sqrt{1 - 4\frac{m_0^2}{A_\lambda^2}}$, $<H_u>\,=\,<H_d>\,=0$) is a minimum of our potential.  It was noted that this is not the global minimum.  In fact this is to be found at
\begin{align}
  <H>\,=\pm\frac{A_\lambda}{2\lambda}\sqrt{1-\frac{4m_0^2}{A_\lambda^2}}
\end{align}
\begin{align}
  <\phi>\,=\frac{-A_\lambda}{2\lambda}
\end{align}
\begin{align}
  <N>\,=0.
\end{align}

While the existence of this ``bad'' solution is clearly a drawback of the model it remains physically viable if the transition probability from the local minimum to the global minimum is longer than the age of the universe \cite{Chung:2003fi}.  We also note that, in the case of inverted hybrid inflation, the trajectory is such that the ``good'' minimum is reached first, as discussed in section \ref{eq:Inflation}.

It is worth mentioning that the model could be altered such that the global minimum arises for $N\neq 0$ and $H_u=H_d=0$.  Specifically we could relax the assumptions that $A_\lambda=A_\kappa$ and $\kappa=\lambda$.  If we examine the potentials at both minima we see that
\begin{equation}
  V_{N\neq 0}= V(0) - \frac{A_\kappa^4}{64\kappa^2}\left(1-\frac{4m_0^2}{A_\kappa^2}\right)^2
\end{equation}
and
\begin{equation}
  V_{H\neq 0}= V(0) - \frac{A_\lambda^4}{16\lambda^2}\left(1-\frac{4m_0^2}{A_\lambda^2}\right)^2.
\end{equation}
From these equations we see that if we make $A_\kappa^2/\kappa\gg A_\lambda^2/\lambda$ then $V_{N\neq 0}$ will be promoted to the global minimum.  However doing so increases the complexity of the model and loses touch with the string construction presented in \cite{companion}.

\end{document}